\documentclass{aa}  

\usepackage{graphicx}
\usepackage{amsmath,amsfonts,amssymb}
\usepackage{aas_macros}
\usepackage[usenames]{xcolor}
\usepackage{caption,subfloat}
\usepackage[english]{babel}
\usepackage{gensymb,wasysym}
\usepackage{txfonts}
\usepackage[colorlinks,allcolors=blue]{hyperref}

\newcommand{\figpath}{Figures/}

\begin{document} 

   \title{Planet formation and stability in polar circumbinary discs}
   
   \author{Nicol\'as Cuello
          \inst{1}\fnmsep\inst{2}
          and Cristian A. Giuppone
          \inst{3}
          }

   \institute{Instituto de Astrof\'isica, Pontificia Universidad Cat\'olica de Chile, Santiago, Chile, \email{ncuello@astro.puc.cl}
    \and
    N\'ucleo Milenio de Formaci\'on Planetaria (NPF), Chile,
    \and
    Universidad Nacional de C\'ordoba, Observatorio Astron\'omico - IATE, Laprida 854, X5000BGR C\'ordoba, Argentina
    }

  \authorrunning{N. Cuello \& C.A. Giuppone}

   \date{Received 30 July 2018 / Accepted 25 June 2019}

% \abstract{}{}{}{}{} 
% 5 {} token are mandatory
  \abstract
   {Dynamical studies suggest that most circumbinary discs (CBDs) should be coplanar (i.e. the rotation vectors of the binary and the disc should be aligned). However, some theoretical works show that under certain conditions a CBD can become polar, which means that its rotation vector is orthogonal with respect to the binary orbital plane. Interestingly, very recent observations show that polar CBDs exist in nature (e.g. HD~98800).}
   {We test the predictions of CBD alignment around eccentric binaries based on linear theory. In particular, we compare prograde and retrograde CBD configurations. Then, assuming planets form in these systems, we thoroughly characterise the orbital behaviour and stability of misaligned (P-type) particles. This is done for massless and massive particles.}
   {The evolution of the CBD alignment for various configurations was modelled through three-dimensional hydrodynamical simulations. For the orbital characterisation and the analysis stability, we relied on long-term N-body integrations and structure and chaos indicators, such as $\Delta e$ and {\sc megno}.} 
   {We confirm previous analytical predictions on CBD alignment, but find an unexpected symmetry breaking between prograde and retrograde configurations. More specifically, we observe polar alignment for a retrograde misaligned CBD that was expected to become coplanar with respect to the binary disc plane. Therefore, the likelihood of becoming polar for a highly misaligned CBD is higher than previously thought. Regarding the stability of circumbinary P-type planets (also know as Tatooines), polar orbits are stable over a wide range of binary parameters. In particular, for binary eccentricities below 0.4 the orbits are stable for any value of the binary mass ratio. In the absence of gas, planets with masses below $10^{-5}\,M_{\odot}$ have negligible effects on the binary orbit. Finally, we suggest that mildly eccentric equal-mass binaries should be searched for polar Tatooines.}
  {}

   \maketitle

%%%%%%%%%%%%%%%%%%%%%%%%%%%%%%%%%%%%%%%%

\section{Introduction}
\label{sec:intro}

Planets appear as common by-products of the star formation process within molecular clouds \citep{Chiang&Youdin2010}. At early evolutionary stages (class 0/I), the fraction of stellar binaries and higher-order multiple systems is remarkably high: between 30\% and 70\% \citep{Connelley+2008, Chen+2013}. Consequently, protoplanetary discs can be severely affected by binary formation and stellar flybys (e.g. \citealt{Bate2018,Pfalzner2013}). Given the stochasticity of these processes, highly asymmetrical discs naturally form on timescales of the order of a few hundred  kyr. Therefore, planet formation is also expected to occur in a broad variety of misaligned discs \citep{Bate2018,Cuello+2019b}. Currently, about a hundred planets have been detected in multiple-star systems \citep{Martin2018}. There is, however, a striking tension with the more than 3\,700 planets detected\footnote{www.exoplanet.eu} around single stars \citep{Batalha+2013}.

Planets in binary systems can either be circumbinary (P-type, also known as Tatooines) or circumstellar (S-type). The seeming lack of P-type planets has been ascribed to observational biases and dynamical processes. On the one hand, these systems are intrinsically difficult to observe through radial velocities \citep{Eggenberger&Udry2007, Wright+2012} and transit methods \citep{Martin&Triaud2014}. For transits in non-coplanar systems, the arbitrary orientation of the planet chord coupled to the fact that most binaries do not eclipse renders P-type planet detection extremely challenging. On the other hand, recent stellar-tidal evolution models of short-period binaries show that the binary orbital period increases with time \citep{Fleming+2018}. This translates into a larger region of dynamical instability around the binary, which could explain the lower frequency of P-type planets compared to S-type planets. Based on the available data, typical planets found around binary stellar systems have a radius of the order of $10$ Earth radii and orbital periods of about $160$ days \citep{Martin2018}. For instance, around a binary system with total mass equal to $1 \, M_{\odot}$, this corresponds to an orbit with a semi-major axis of approximately $0.35$ au.

Previous theoretical studies suggested that the disc-binary interaction, namely the gravitational torque, should align the disc with the binary orbital plane on timescales shorter than the disc lifespan \citep{Foucart&Lai2013}. Therefore, until recently, misaligned circumbinary discs (CBDs) were considered exotic. However, the observations of highly non-coplanar systems such as 99~Herculis \citep{Kennedy+2012a}, IRS~43 \citep{Brinch+2016}, GG~Tau \citep{Cazzoletti+2017}, and HD~142527 \citep{Avenhaus+2017} suggest otherwise. For example, in the multiple system of GG~Tau, \cite{Aly+2018} recently explored various initial conditions to explain the disc misalignment with respect to the binary. Even more recently, \cite{Kennedy+2019} reported the very first observation of a circumbinary gas-rich disc in a polar configuration. These discoveries are rather puzzling and deserve to be explained.

In the field of secular dynamics, the seminal work by \cite{Ziglin1975} presented an integrable approximation for the orbits of planets around binary systems. Later, based on numerical calculations of circumbinary polar particles, \citet{Verrier+2009} reported the coupling between the inclination and the node. Then, \cite{Farago&Laskar2010} studied the circumbinary elliptical restricted three-body problem, giving an averaged quadrupolar Hamiltonian. In this way the authors showed that equilibrium configurations can be reached at high inclinations in the restricted and massive three-body problem. Then, relying on this analysis, it is possible to define the separatrix around the equilibrium points.

A more thorough analysis of the stability of massless particles was done by \cite{Doolin+2011} where they consider different binary eccentricities and mass ratios between the components. This approach allows us to unambiguously identify prograde, retrograde, and polar orbits. In addition, the authors study the validity of the quadrupolar Hamiltonian approximation when the distance to the binary centre of mass is not sufficiently small. Furthermore, based on the study of hierarchical triple systems, several other studies investigated the circumbinary polar orbits for the restricted problem considering the octupole Hamiltonian \citep{Ford+2000, Naoz+2013, Li+2014}. Interestingly, for near polar configurations, the inner binary can significantly excite the orbital eccentricity (up to about 0.3 in some cases). This excitation is weakly dependent on the distance to the binary. More recently, \cite{Naoz+2017} and \cite{Zanardi+2018} studied the evolution of polar orbits considering additional effects such as general relativity terms. Last but not least, the effect of the binary on the CBD can have dramatic effects on the disc alignment as described below.

Given the recent works on polar alignment and CBD dynamics (Sect.~\ref{sec:previous}), it is interesting to explore further the conditions under which polar alignment of misaligned circumbinary discs occurs (Sect.~\ref{sec:CBDs}). Then, assuming planet formation happens in these discs, we characterise the orbits of single P-type planets around eccentric and inclined binaries (Sect.~\ref{sec:characterisation}). This allows us to constrain the binary eccentricity and mass ratio of systems around which we expect to observe polar circumbinary planets in the future. We discuss our findings in Sect.~\ref{sec:discussion} and finally conclude in Sect.~\ref{sec:conclusions}.

%%%%%%%%%%%%%%%%%%%%%%%%%%%%%%%%%%%%%%%%

\section{Summary of previous works}
\label{sec:previous}

In the following we list the series of works that have previously addressed the topic of misaligned CBDs. We pay particular attention to nodal libration, polar CBDs, and alignment conditions.
    
\subsection{Nodal libration, binary parameters, and stability}
\label{sec:bin}

\cite{Doolin+2011} numerically investigated the nodal libration mechanism in the longitude of the ascending node and in the inclination for P-type particles around eccentric binaries. This phenomenon occurs for highly inclined orbits and for a wide range of binary mass ratios. Polar orbits were found to be remarkably stable at distances of two to three times the binary semi-major axis, which is not the case for coplanar configurations. Moreover, between the islands of stability, the authors report vertical striations of instability and hypothesise that they might be due to orbital resonances (see their Fig.~14). More recently, \cite{Zanazzi&Lai2018} derived a simple analytical criterion based on the binary eccentricity and initial disc orientation that assesses whether the inclination of the CBD is expected to librate or not. It is worth noting that this criterion is symmetric with respect to the disc orientation: it does not distinguish prograde from retrograde configurations. However, as we show in Sect.~\ref{sec:broken}, this condition is not sufficient in case of symmetry breaking.

\subsection{Polar alignment of CBDs}
\label{sec:polaralignment}

In \cite{Aly+2015} and \cite{Martin&Lubow2017} it is demonstrated through numerical simulations how inclined gaseous CBDs around eccentric binaries can tidally evolve towards polar configuration with respect to the binary orbital plane. In this case, the dissipative torque acting on the disc is key to damping the nodal libration. This mechanism is investigated further in \cite{Zanazzi&Lai2018} and \cite{Lubow&Martin2018} where an analytical framework is provided to describe the polar alignment of CBDs, taking into account the gaseous viscosity of the disc.

\subsection{Flybys and generation of polar alignment conditions}
\label{sec:flyby}

Stellar flybys, also known as tidal encounters, provide a natural way to tilt and twist a CBD. More specifically, a disc initially aligned with the binary orbital plane can be tilted by several tens of degrees and twisted by roughly 90$\degr$ after a flyby of a massive star \citep[figure D6]{Cuello+2019b}. In particular, the smaller the pericentre distance and the more massive the perturber, the higher the resulting disc misalignment. It is worth noting that this flyby must happen in the early stages of disc formation such that enough time should remain for the disc to reach the polar configuration. In this scenario, inclined and retrograde encounters are the most efficient for disc warping (\citealt{Xiang-Gruess2016}). Alternatively, misaligned accretion onto the binary, as in \cite{Bate2018}, provides another mechanism to reach appropriate conditions for polar alignment.

\subsection{CBD parameters and alignment timescales}
\label{sec:timescales}

More recently, \citet[hereafter ML18]{Martin&Lubow2018} explored the alignment of CBDs in a broad variety of configurations through hydrodynamical simulations. Specifically, they studied the dependence of the disc viscosity, temperature, size, binary mass ratio, orbital eccentricity, and inclination on the disc evolution. The authors find that polar alignment occurs for a wide range of binary--disc parameters, especially for low mass discs. Their results broadly agree with the expectation of linear theory. Interestingly, for large binary separations and/or for low mass ratios, the alignment timescale might be longer than the expected circumbinary disc lifetime. If this is the case, the CBD might remain on a misaligned configuration, neither (anti-)aligned nor polar. This suggests that highly inclined discs and planets may exist around eccentric binaries \citep{Zanazzi&Lai2018}.

%%%%%%%%%%%%%%%%%%%%%%%%%%%%%%%%%%%%%%%%

\section{Circumbinary disc dynamics and polar alignment}
\label{sec:CBDs}

We first summarise the conditions under which we expect polar alignment. Then we present our SPH simulations, and finally describe the unexpected symmetry breaking observed.

\subsection{Binary parameters and polar orbits}
\label{binaryparams}

\begin{figure}
  \centering
\mbox{\includegraphics[width=0.38\textwidth]{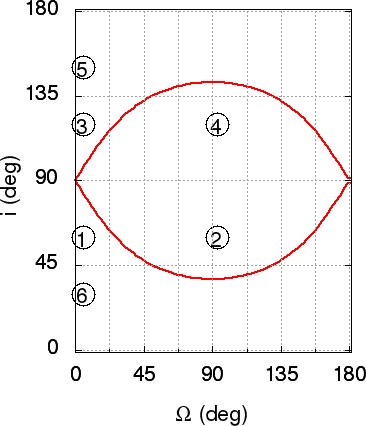}} 
 \caption{Separatrix in the ($i$, $\Omega$) plane constructed from Eq.~\ref{eq:F} for $e_B=0.5$. Initial conditions inside the separatrix limits correspond to polar orbits 2 and 4, whereas those outside correspond to prograde ($i<90\degr$: 1, 6) and retrograde ($i>90\degr$: 3, 5). The initial conditions of the SPH simulations of Table~\ref{tab:sims} are identified with numbers.}
  \label{fig:eb}
\end{figure}

We consider a binary system with total mass $M$ and individual masses $M_1$ and $M_2$. We call $q=M_2/M_1$ the binary mass ratio. To describe the motion of a P-type particle around the binary we use the following Jacobi orbital elements: semi-major axis $a$, eccentricity $e$, inclination $i$ (with respect to the binary orbital plane), mean longitude $\lambda$, longitude of pericentre $\varpi$ (alternatively argument of pericentre $\omega$), and longitude of the ascending node $\Omega$. The sub-index $B$ is used when referring to the binary orbit. Without loss of generality, we set initial conditions \{$\lambda_{\rm B}=0\degr$, $\varpi_{\rm B}=0\degr$, $\Omega_{\rm B}=0\degr$\}. The angles of a given particle are measured from the direction of the binary pericentre. Finally, we set our fiducial system $M_1=M_2=0.5 \, M_\odot$ (i.e. $q=1$), $a_{\rm B} = 0.1$ au, and $e_{\rm B}=0.5$. This allows us to directly compare our results with those in \cite{Martin&Lubow2017}.
Regarding the dynamics of P-type particles, the phase space depends on $\Omega$ with two equilibrium points at \{$\Omega=\pm90\degr, \, i=90\degr$\}, where both angles librate \citep[see][for a detailed analysis]{Doolin+2011}. Interestingly, for any given couple of $(\Omega,\,i)$, it is possible to assert whether $i$ librates around $\pm90^\circ$ or not by calculating the separatrix $F$ \citep{Farago&Laskar2010}:
\begin{equation}
\label{eq:F}
F = \sqrt{\frac{1-e_B^2}{1-5e_B^2\cos^2\Omega+4e_B^2}} \,\,\,.
\end{equation}
Specifically, its orbit is called polar if a particle fulfils the following condition:
\begin{equation}\label{eq:limit}
    \arcsin F < i < \pi - \arcsin F \,\,\,.
\end{equation}
 In Figure~\ref{fig:eb} we show the separatrix in the plane ($\Omega,\,i$) for the fiducial parameters considered. The initial conditions inside the separatrix limits correspond to the region of polar orbits for this given configuration. We recall, as described in Sect.~\ref{sec:polaralignment}, that the higher the value of $e_{\rm B}$, the larger the  region.

To sum up, in order to be polar, a P-type particle needs to ~have a longitude of the ascending node close to $90\degr$ and ~be orbiting around an eccentric binary on a highly inclined orbit (see Fig.~\ref{fig:eb}). When these conditions are met and in the absence of gas, then the particle's inclination and the ascending node librate around $90\degr$.

\subsection{Misaligned circumbinary disc simulations}
\label{sec:hydro}

In this section, we assume that the chaotic and violent stellar cluster environment has produced an initially tilted CBD (see Sect.~\ref{sec:flyby}). We computed its resulting alignment, considering different initial disc inclinations ($i_0$) and twists ($\Omega_0$). For simplicity, we only considered the set of fiducial parameters for the binary.

We ran 3D hydrodynamical simulations of CBDs by means of the {\sc phantom} Smoothed Particle Hydrodynamics (SPH) code \citep{Price+2018}. The SPH method (see e.g. \citealt{Price2012}) is well suited for misaligned CBDs simulations because there is no preferred geometry and angular momentum is conserved to the accuracy of the time-stepping scheme, irrespective of the plane of the binary orbit. Both stars are represented by sink particles with accretion radii equal to $0.25\,a_{\rm B}$, which interact with the gas via gravity and accretion \citep{Bate+1995}. We used $10^6$ gas particles to model the disc, which is a resolution high enough to avoid artificially high numerical viscosities\footnote{see ML18 for a detailed resolution study of this kind of simulation.}. The inner and outer radii of the disc were set to $R_{\rm in}=2\,a_{\rm B}$ and $R_{\rm out}=5\,a_{\rm B}$, respectively. The surface density was initially set to a power law profile equal to $\Sigma \propto R^{-3/2}$. The disc total mass is equal to $0.001 \, M_{\odot}$. Given this low value, we neglected the disc self-gravity in our calculations. Furthermore, we assumed that the disc is locally isothermal (i.e. that the sound speed follows $c_{\rm s} \propto r^{-3/4}$ and $H/R=0.1$ at $R=R_{\rm in}$). For further details on the {\sc phantom} disc set-up (see \citealt{Price+2018}). Finally, we adopted a mean Shakura-Sunyaev disc viscosity $\alpha_{\rm SS} \approx 0.01$.

\begin{table}
\begin{center}
\caption{Set of circumbinary disc simulations for different initial tilts ($i_0$) and twists ($\Omega_0$). As in ML18, the disc either circulates (C) or librates (L).}
\label{tab:sims}
\begin{tabular}{|c|c|c|c|c|c|}
\hline
Run & $i_0$ ($\degree$) & $\Omega_0$ ($\degree$) & $i_{\infty}$ ($\degree$) & C or L \\
\hline
    r1 & 60 & 0 & 0 & C \\
    r2 & 60 & 90 & 90 & L \\
    r3 & 120 & 0 & 90 & L \\
    r4 & 120 & 90 & 90 & L \\
    r5 & 150 & 0 & 0 & C \\
    r6 & 30 & 0 & 0 & C \\
 \hline
\end{tabular}
\vspace{-1.5em}
\end{center}
\end{table}

\begin{figure}
\begin{center}
\includegraphics[width=0.43\textwidth]{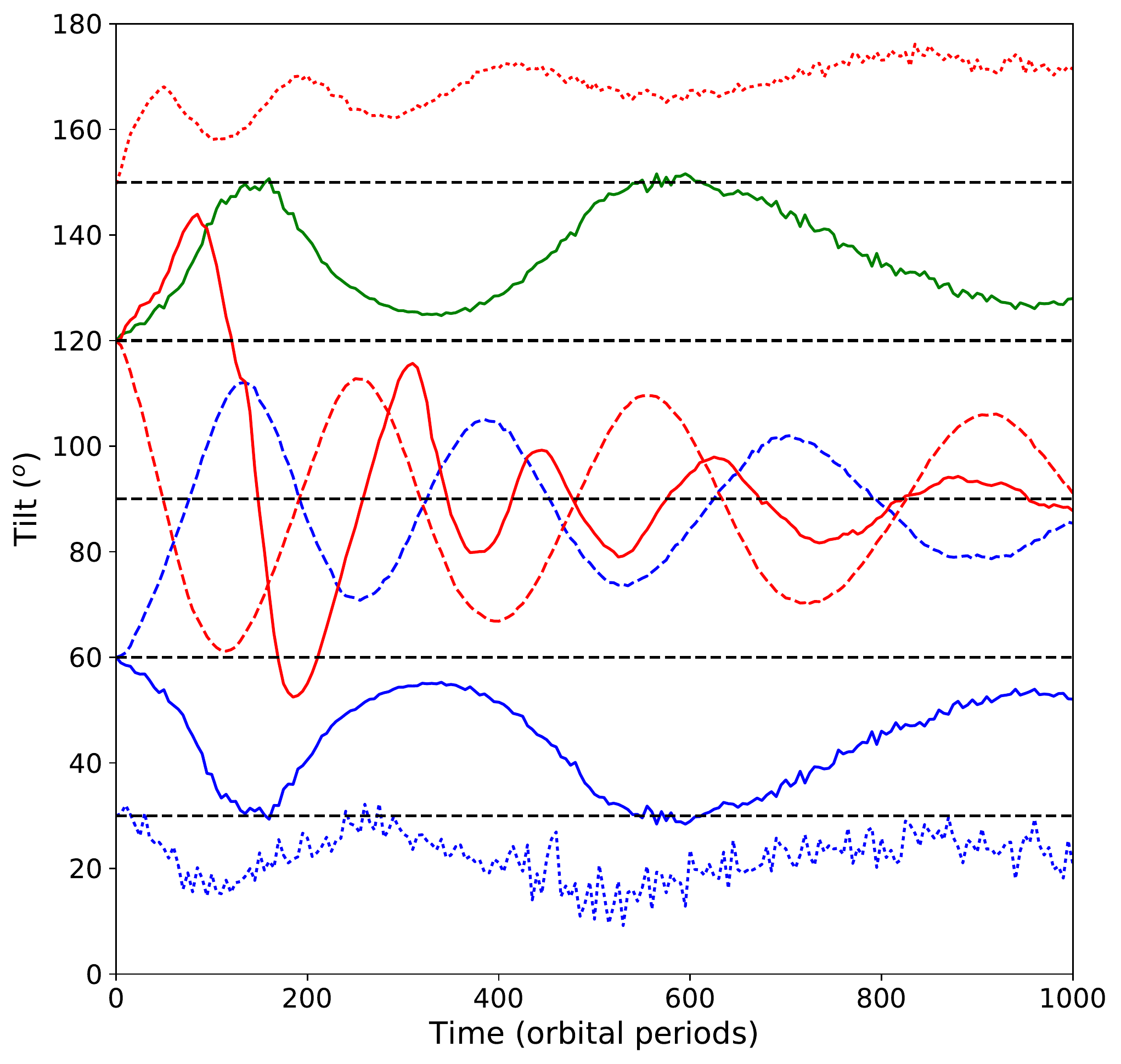}
\caption{Tilt evolution over time, $i(t)$, in terms of binary orbital periods. Runs: r1 (solid blue), r2 (dashed blue), r3 (solid red), r4 (dashed red), r5 (dotted red), r6 (dotted blue). For comparison with r3, we also plot the symmetric curve of r1 with respect to $i~=~90\degr$ (solid green). r3 shows an anomalous behaviour because the disc breaks in two discs after $\sim 150$ orbits, which then evolve towards polar alignment} (see Sect.~\ref{sec:broken}).
\label{fig:tiltvst}
\vspace{-1.5em}
\end{center}
\end{figure}

Our set of {\sc phantom} simulations is given in Table~\ref{tab:sims}. We only change the disc inclination ($i_0$) and its argument of ascending node ($\Omega_0$). For completeness, these disc initial conditions are plotted as numbered dots in Fig.~\ref{fig:eb}. Based on the linear theory in \cite{Zanazzi&Lai2018}, the disc is expected to become polar in 2 and 4, align in 1 and 6, and anti-align in 3 and 5, with respect to the binary plane. From the practical point of view, we decompose the disc in 500 annuli and compute the tilt and the twist for each one of them. There is little difference between the disc warping in the inner and in the outer regions, except for r3 (see Sect.~\ref{sec:broken}). Therefore, unless stated otherwise, we compute the tilt ($i$) and the twist ($\Omega$) at $r= 3 \, a_{\rm B}$ from the {\sc phantom} outputs.

\begin{figure*}
\begin{center}
\includegraphics[width=1.0\textwidth]{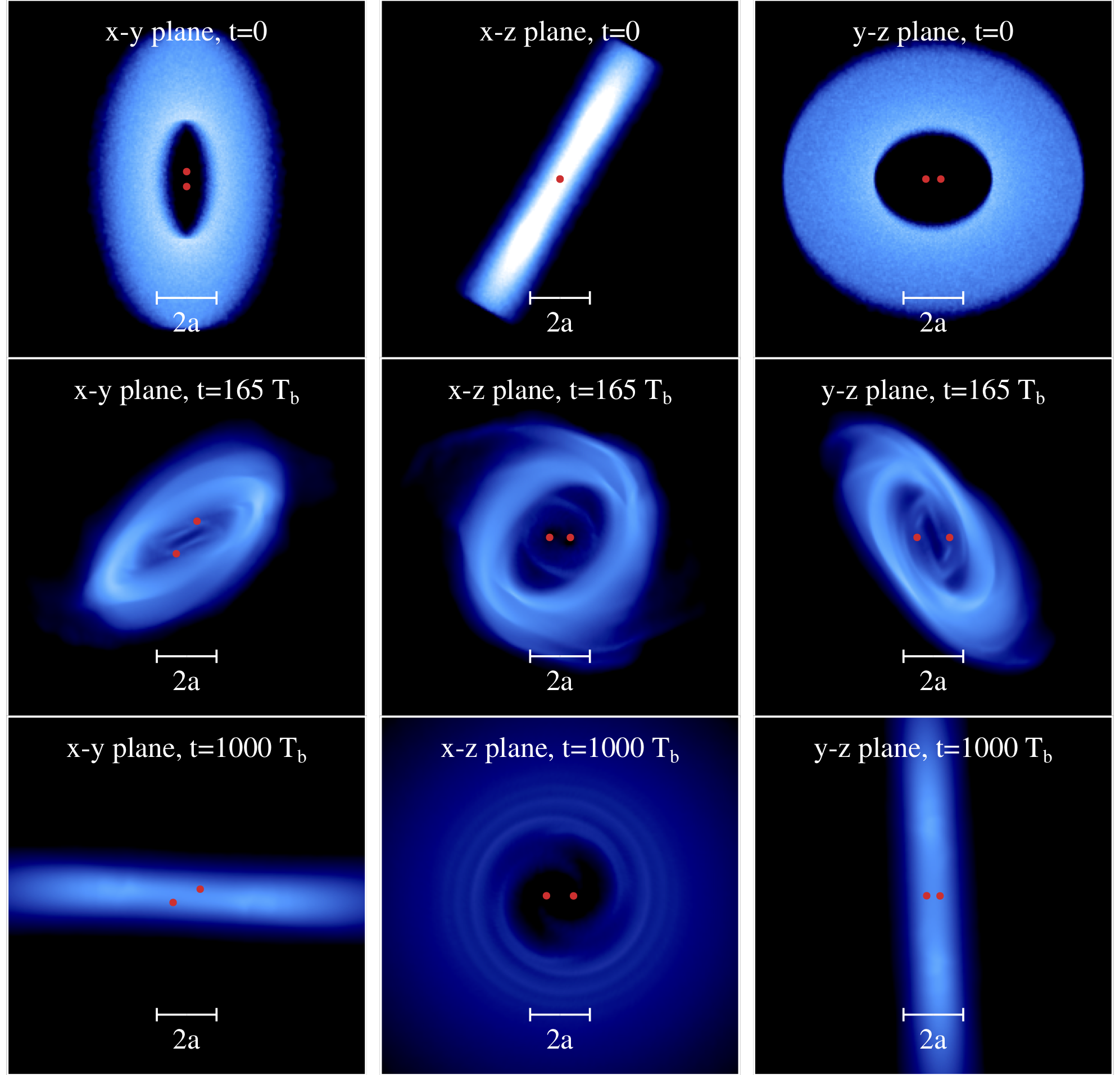}
\caption{Circumbinary disc in r3: the binary plane lies in the $xy$-plane and the disc is initially inclined ($i_0=120\degr$, $\Omega_0=0\degr$). At 165 binary orbits, we observe that the tilt is different between the inner and the outer regions of the disc (cf. Fig.~\ref{fig:r3tilt}). This is interpreted as the disc being broken in two discs by the binary. The inner regions experience polar alignment, causing the whole disc to become polar after roughly $400$ binary orbits (cf. Fig.~\ref{fig:tiltvst}). At $1\,000$ binary orbits, the disc is almost perfectly polar.}
\label{fig:r3-xyz}
%\vspace{-1.5em}
\end{center}
\end{figure*}

In Figure~\ref{fig:tiltvst} we plot the evolution of the tilt over time for $1\,000$ binary orbital periods. We obtain the expected alignment for all the runs but r3. On the one hand, in r1 and r6 the disc inclination remains prograde and is slowly damped, while $\Omega$ circulates (not shown). In r5 the disc anti-aligns with respect to the binary ($i=180\degr$), while $\Omega$ circulates. On the other hand, in r2, r3, and r4 the disc becomes polar ($i=90\degr$) and $\Omega$ librates (instead of circulating). Remarkably, r2 and r4 are in phase opposition and the oscillations are quickly damped over time (compared to r1 and r6). Also, for low inclinations (r5 and r6) the disc becomes coplanar more quickly in the retrograde configuration (r5) compared to the prograde (r6) because the inner cavity, opened by resonant torques, is smaller for retrograde configurations \citep{Miranda&Lai2015, Nixon&Lubow2015}. Therefore, the disc responds more quickly to the binary torques. The anomalous behaviour of r3 is detailed below.

\subsection{Unexpected symmetry breaking in r3} \label{sec:broken}

\begin{figure}
\begin{center}
\includegraphics[width=0.5\textwidth]{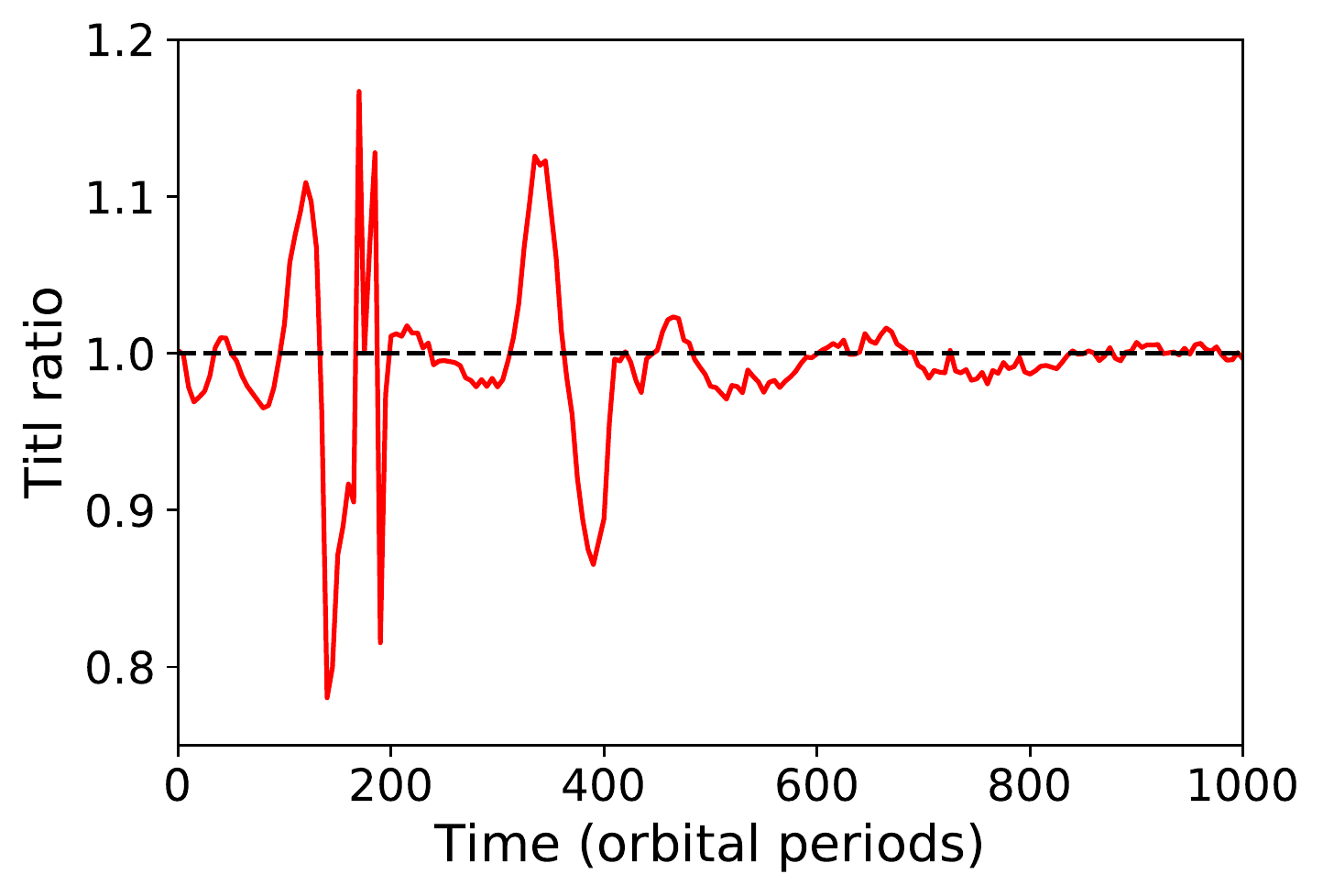}
\caption{Tilt ratio between $a=5\,a_{\rm B}$ and $a=2\,a_{\rm B}$ in r3. When the disc behaves as a solid body, this ratio is $\sim 1$. The oscillations of this ratio (between 0.8 and 1.2) appear when the disc breaks in two misaligned discs. This occurs between approximately $150$ and $400$ binary orbits (see middle row in Fig.~\ref{fig:r3-xyz}).}
\label{fig:r3tilt}
\end{center}
\end{figure}

In Fig.~\ref{fig:r3-xyz} we show the CBD in r3 for three evolutionary stages: 0, 165, and 1000 binary orbital periods. Given the initial parameters of r3, the disc should anti-align with respect to the binary (see Fig.~\ref{fig:eb}). However, after half a tilt oscillation ($\sim150$ binary periods), the disc suddenly fulfils the condition for polar alignment and its tilt starts to oscillate around $90\degr$. Additionally, its twist starts to librate instead of circulating. Both oscillations are damped, as expected for the viscous evolution of a gaseous disc (ML18). Hence, this shows that the symmetry between prograde and retrograde orbits is broken in some specific cases (e.g. r3).

In Figure~\ref{fig:r3tilt}, we show the evolution of the ratio between the outer ($a=5 \, a_{\rm B}$) and the inner ($a=2 \, a_{\rm B}$) regions of the circumbinary disc in r3. This quantity indicates whether the disc is behaving as a solid body ($\sim 1$) or not. By construction, this ratio is initially equal to $1$. However, after a hundred binary orbits, it starts to oscillate between $0.8$ and $1.2$ due to the disc breaking, which is caused by the binary torque (see middle row in Fig.~\ref{fig:r3-xyz}). We observe that the inner disc regions undergo a fast polar alignment, which eventually causes the entire CBD to align at $i=90\degr$. After $400$ binary orbits the entire disc has reached polar alignment and behaves as a solid body again (i.e. the tilt ratio is close to $1$). This phenomenon is representative of the symmetry breaking between prograde and retrograde configurations.

The polar alignment in r3 is anomalous in the sense that it cannot be predicted by the analytical theory. More specifically, the expression of $\Lambda$ in Eq.~7 in \cite{Zanazzi&Lai2018}, which shows whether the CBDs is expected to become polar or not, is symmetric for prograde and retrograde orbits (see e.g. their figure~ 2). Therefore, fulfilling the polar alignment criterion at the very beginning is not a necessary condition for polar alignment. The cavity size and further non-linear effects play a key role in determining the final CBD alignment. Finally, as in ML18, we find that during the simulations the variations in the binary orbital parameters are only of the order of a few percent.

Polar CBDs might then be more common than previously thought. Therefore, since planet formation is expected to occur within these misaligned discs, it is relevant to characterise the planetary orbits around eccentric and inclined binaries.

%%%%%%%%%%%%%%%%%%%%%%%%%%%%%%%%%%%%%%%%

\section{P-type planet orbital characterisation}
\label{sec:characterisation}

\begin{figure*}
  \centering
\mbox{\includegraphics[width=0.71\textwidth]{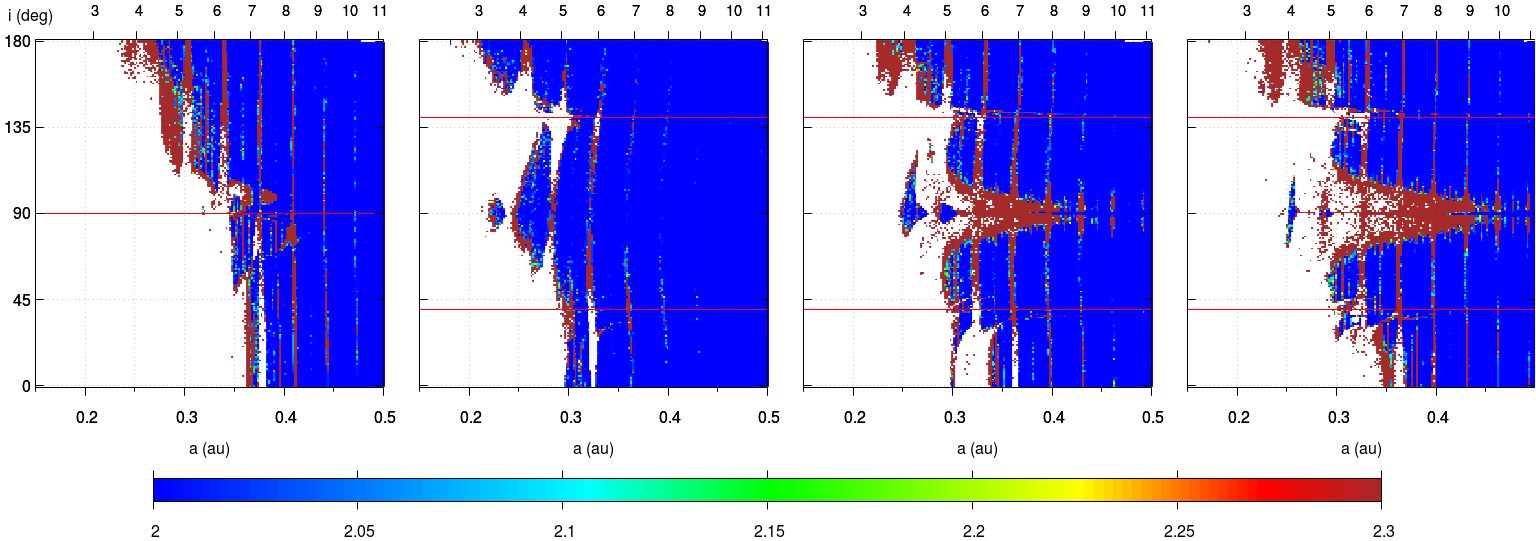}} \mbox{\includegraphics[width=0.71\textwidth]{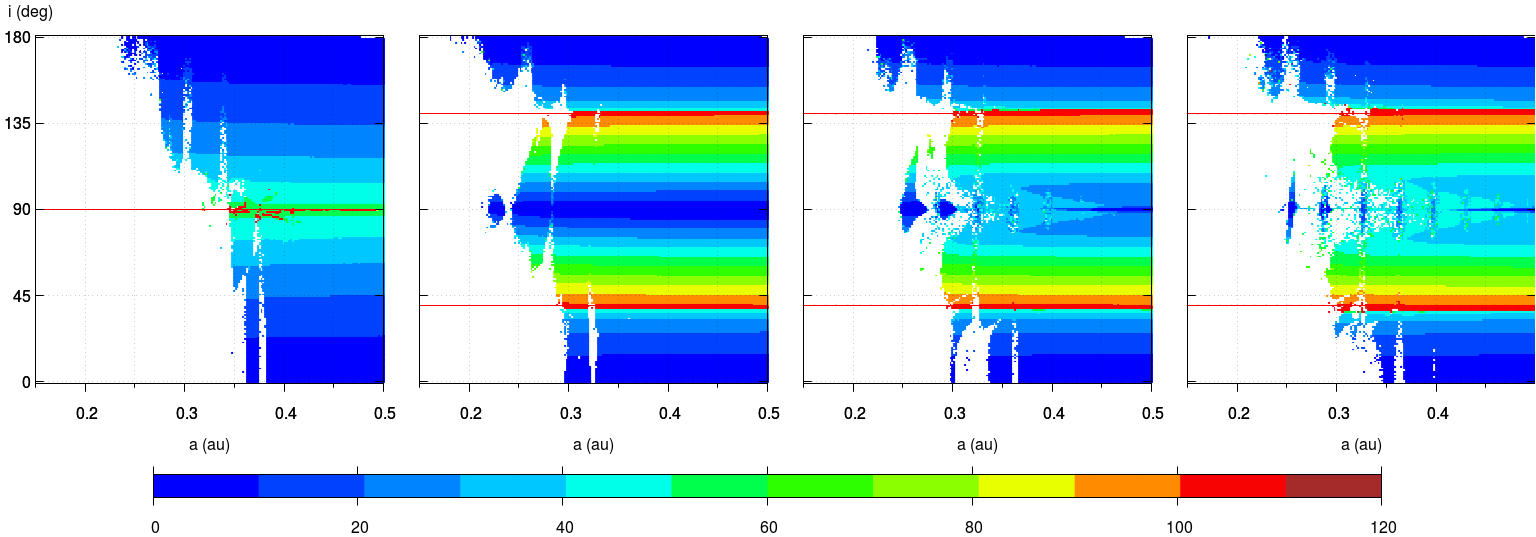}}
\caption{Dynamical maps of inclined systems with $M=1\,M_{\odot}$ (from left to right): \{$q=1$, $\Omega=0\degr$\}, \{$q=1$, $\Omega=90\degr$\}, \{$q=0.5$, $\Omega=90\degr$\}, and \{$q=0.2$, $\Omega=90\degr$\}. The top labels correspond to the period in terms of binary periods. The regions between the horizontal red lines correspond to real polar orbits (i.e. librating around $i=90\degr$). Long-term integrations show that particles with initial conditions in the red/brown regions do not survive more than $10^5$ periods. When $\Omega=90\degr$, stability islands appear at polar inclinations closer to the binary (between $2$ and $3 \, a_{\rm B}$). In the bottom panel, the colour scale shows the amplitude of libration of $i$. The continuous red lines are defined by Eq.~\eqref{eq:limit} and correspond to the analytic limits between polar and prograde or retrograde orbits.}
\label{fig:ai}
\end{figure*}

\begin{figure*}
\begin{center}
\includegraphics[width=0.95\textwidth]{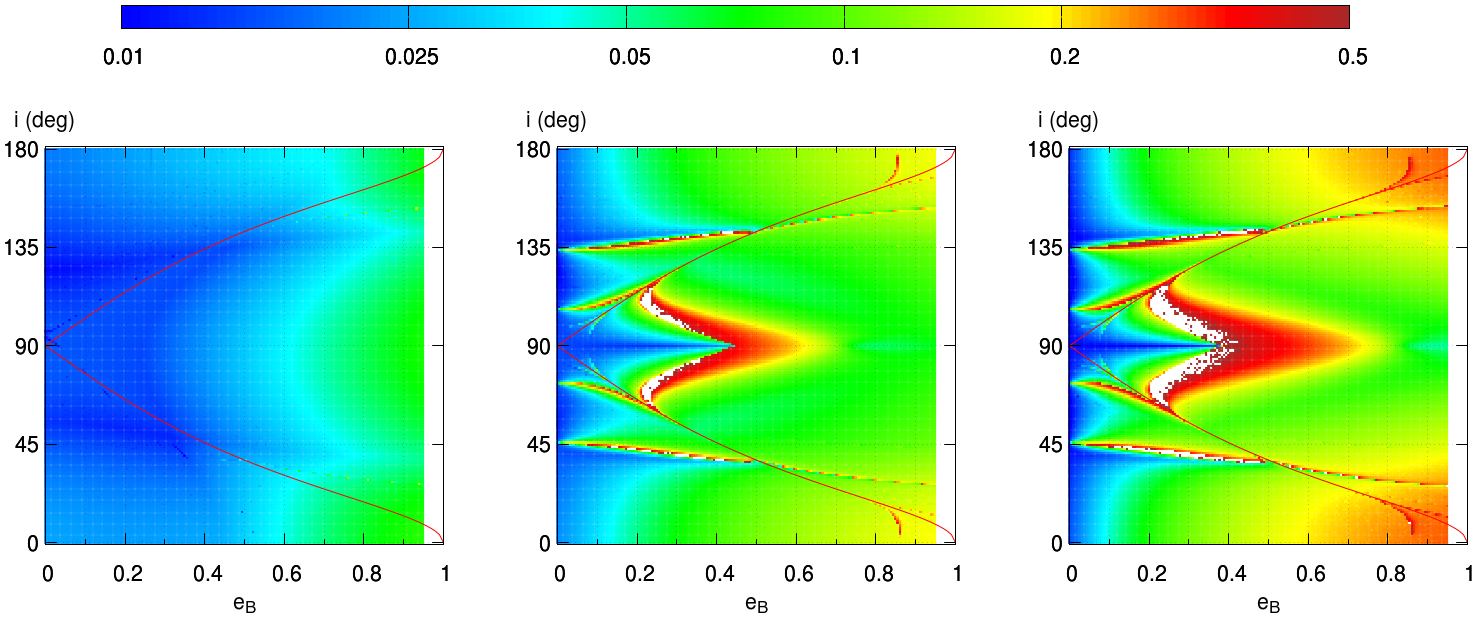}\\  \vspace{-1.0em}
\includegraphics[width=0.95\textwidth]{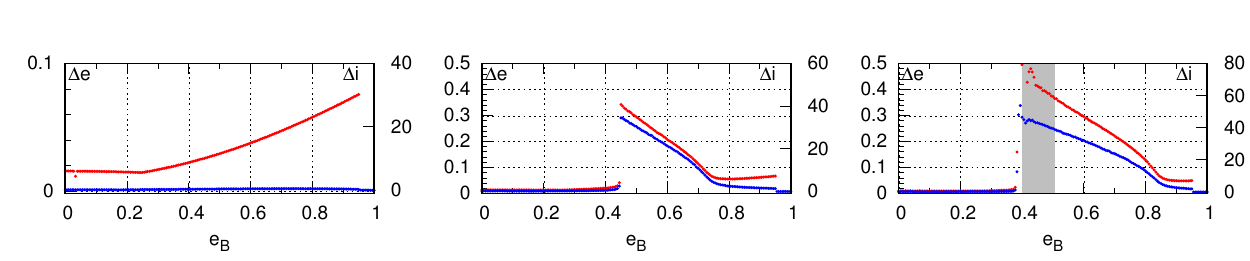}\\
\caption{Top frames: Dynamical maps of inclined configurations with $\Omega=90\degr$ with q=$1,\,0.5,\,0.2$ (from left to right). {The colour scale corresponds to the eccentricity variation $\Delta e$ of the orbit.} The continuous red lines are defined by Eq.~\ref{eq:F} and correspond to the analytic limit between polar and prograde or retrograde orbits. Bottom frames: {Longitudinal cuts of the dynamical maps at $i=90^\circ$}. The amplitude of $\Delta e$ ($\Delta i$) is shown in red (blue). The grey shaded areas correspond to chaotic orbits with $\langle Y \rangle \gg 2$.}
\label{fig:q}
\vspace{-1.5em}
\end{center}
\end{figure*}

We focus on the stability and evolution of single circumbinary P-type planets ($a > a_{\rm B}$) for a wide variety of binaries. To do so, we construct a regular 2D mesh of initial conditions with a pair of orbital parameters (e.g. $i$ and $e$), while the remaining four parameters (e.g. $a$, $\lambda$, $\omega$, and $\Omega$) are initially set at nominal values. Then, we solve the three-body equations of motion numerically by means of a double-precision Bulirsch-Stoer integrator with tolerance $10^{-12}$.

\subsection{Dynamics of an inclined massless planet}
\label{sec:massless}

First, we focus on the case of massless particles. A thorough study of the survival times for a regular combination of binary eccentricity $0<e_B<0.9$ and mass ratio $ 0.2<q<1$ in the ($a,\,i$) plane can be found in \cite{Doolin+2011}. Here, we extend their analysis by focusing on the features observed in the dynamical maps in order to better characterise inclined orbits. In particular, we use further structure indicators.

In Figure \ref{fig:ai}, we show the dynamical maps in the ($a,\,i$) plane for $\Omega=0\degr$ (leftmost panels) and $\Omega=90\degr$ (rightmost panels). Each orbit is integrated for 80\,000 binary periods. If the particle either collides with one of the stars or escapes from the system, then it is coloured in white. In the top row, the colour scale is proportional to the Mean Exponential Growth of Nearby Orbits ({\sc megno}) value $\langle Y \rangle$. This indicator is particularly appropriate because it identifies the chaotic orbits at a low CPU-cost \citep{Cincotta&Simo2000}. Regular orbits have $\langle Y \rangle \sim 2$ (blue), while chaotic and potentially unstable orbits have $\langle Y \rangle > 2$ (red/brown). In the bottom row, the colour scales correspond to the amplitude variation of $i$.

On the one hand, when $\Omega=0\degr$ (first column in Fig.~\ref{fig:ai}), the prograde orbits are stable only beyond $3.5 \, a_{\rm B}$, while retrograde orbits are stable closer to the binary system ($a \sim 2.2 \, a_{\rm B}$). This occurs because the overlap of nearby resonances for retrograde orbits is of higher order compared to the case of prograde orbits \citep[see][]{Morais_2012}. The vertical red stripes are associated with the $1:N$ mean motion resonances (MMR), where eccentricity is efficiently excited. The orbits of these region have $\langle Y \rangle \gg 2$, which indicates their chaotic nature. For prograde cases the $1:7$ MMR delimits the stable orbits from the unstable ones at roughly $a\sim 3.5 \, a_{\rm B}$. Alternatively, for retrograde cases this happens for the $1:4$ MMR at $a\sim {2.5} \, a_{\rm B}$. Interestingly, for high initial inclinations the orbits are stable for both prograde and retrograde configurations, even fairly close to $90^\circ$. It is worth noting that this does not occur for S-type configurations \citep[see white gaps in figure 5 in][]{Giuppone+2017}.

On the other hand, when $\Omega=90\degr$ (rightmost columns in Fig.~\ref{fig:q}), the stability maps change significantly and different stable regions appear. Specifically, there are stable orbits with high inclinations ($45\degr \leq i \leq 135\degr$) as close as retrograde orbits. Polar orbits are remarkably stable at small semi-major axis ($a\sim2.5 \, a_{\rm B}$). Furthermore, we analyse the impact of the binary mass ratio $q$ on the stability of the orbits. The two rightmost columns of Figure \ref{fig:ai} show the stability maps for $q=0.5$ and $q=0.2$. As the mass ratio decreases, the MMRs become more important and severely deplete the polar regions ($i\sim90\degr$). These vertical stripes in the $e_{\rm B}$--$q$ parameter space were first reported by \cite{Doolin+2011}. Here, we demonstrate that these peculiar patterns are indeed located at MMRs. In addition for non-equal-mass binaries the libration amplitude of $i$ (and consequently of the nodes) increases for polar orbits (see Fig.~\ref{fig:ai}, bottom frame). This effect correlates well with the chaotic nature of these orbits.

To further characterise the orbits, we construct additional dynamical maps in a regular ($e_{\rm B},\,i$) grid. In Figure~\ref{fig:q}, we show these maps for a fixed semi-major axis equal to $a=4.5 \, a_{\rm B}$ and $\Omega=90\degr$. This corresponds to a regular region of motion according to Fig.~\ref{fig:ai}. Again, this is done for different binary mass ratios: $q=1,\,0.5,$ and $0.2$. To visually identify the region of polar orbits, we overplot a continuous red line in Fig.~\ref{fig:q} defined by Eq.~\eqref{eq:limit}. The regions with the highest eccentricity excitation ($\Delta e \gtrsim 0.5$) or with $e_B \geqslant 0.945$ are unstable (coloured in white). When $q=1$ (leftmost frame) the dynamical maps exhibit low eccentricity amplitudes with {\sc megno} values indicating regular orbits. For lower values of $q$, horizontal stripes of unstable and chaotic orbits appear at $i=i_{\rm N}$ and $i=i_{c}$. The values of $i_{N}\sim73^\circ \text{and } 134^\circ$ correspond to the very narrow octupole resonance wherein $\varpi + \varpi_B - 2 \Omega$
librates about $0^\circ$, while $i_{c}\sim 46^\circ \text{and } 107^\circ$, where the frequency $\dot{\varpi}\sim0$ at $\omega\pm180^\circ$ \citep[see e.g.][for a detailed analysis of these configurations]{Gallardo+2012, Vinson+2018}. As a rule of thumb, the lower the mass ratio $q$, the larger the size of the chaotic region at $i=90^\circ$.

We also observe a diamond-shaped feature centred at $i=90^\circ$ and delimited by the separatrix with a vertex at $e_B\sim0.45$. We checked that the regions where $\Delta e \gtrsim 0.4$ have high values of {\sc megno} (not shown) and are hence chaotic. In the bottom frame of Fig. \ref{fig:q}, we show longitudinal cuts of the dynamical maps at $i=90^\circ$. The curves show the variations in eccentricity ($\Delta e$) and inclination ($\Delta i$) for different values of $e_{\rm B}$. In the leftmost panel ($q=1$) there is no discontinuity observed for $\Delta e$ or $\Delta i$. However, for $q=0.5$, we observe an abrupt change in $\Delta e$ and $\Delta i$ at $e_{\rm B}\sim0.42$. For $q=0.2$, this occurs when $0.4\lesssim e_{\rm B} \lesssim 0.5$. These curves clearly show that $\Delta e$ and $\Delta i$ are strongly correlated.

\subsection{Orbital characterisation for a massless polar planet}

\begin{figure}
\begin{center}
\includegraphics[width=0.45\textwidth]{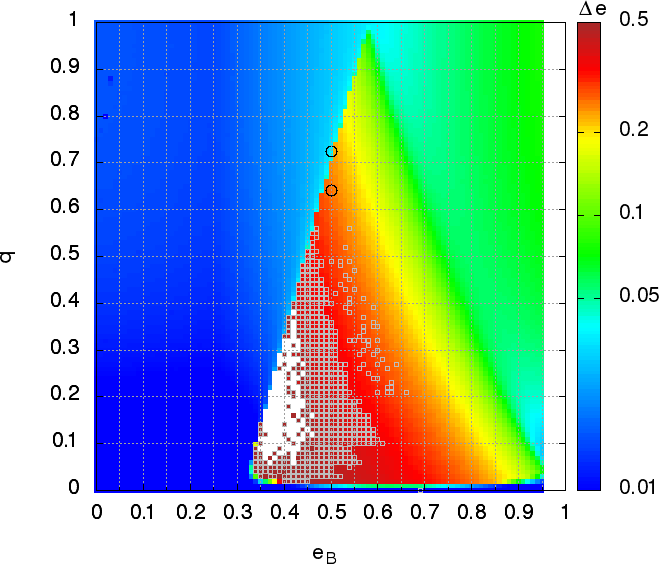}\\
\caption{Dynamical map for polar orbits for massless planets with initial $a=4.5 \, a_B$ in the ($e_B,\,q$) plane. The colour scale corresponds to $\Delta e$. The orbits identified with white dots are unstable. Chaotic orbits identified with the {\sc megno} indicator are indicated with grey squares.}
\label{fig:ebJ}
\vspace{-1.5em}
\end{center}
\end{figure}

In the following, we exclusively focus on polar orbits (i.e. those that are initially at $\Omega=90^\circ,\,i=90^\circ$). Figure \ref{fig:ebJ} shows the polar orbits in the ($e_B,\,q$) plane. The colour scale corresponds to $\Delta e$. High values of $\Delta e \gtrsim 0.3$ result into chaotic behaviour. We see that some of the orbits (white dots) are unstable in less than $10^5$ binary periods. This dynamical map allows us to conclude that binaries with $0.35 \lesssim e_B \lesssim 0.5$ and $0.05 \lesssim q \lesssim 0.4$ are unlikely to harbour polar planets. In other words, this corresponds to the forbidden region for polar type-P circumbinary planets.

\subsection{Orbital characterisation for a massive polar planet}
\label{sec:massive}

\begin{figure}
\begin{center}
\includegraphics[width=0.47\textwidth]{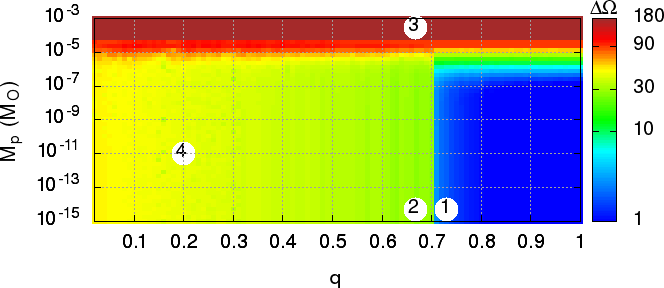}\\
\includegraphics[width=0.46\textwidth]{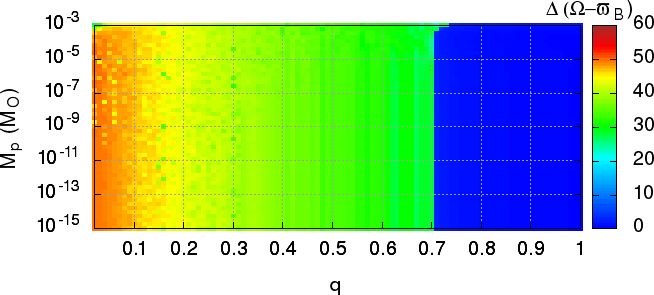}\\
\caption{Dynamical maps of polar orbits ($i=90\degr$, $\Omega=90\degr$) for massive planets in the ($q,M_{\rm p}$) plane with $e_B=0.5$. The colour scale in the upper and lower panels corresponds to $\Delta\Omega$ and $\Delta(\Omega-\varpi_{\rm B})$, respectively. Initially, the planet is located at the equilibrium point $i=90^\circ$, $\Omega=90^\circ$. The four initial conditions identified with white circles correspond to the orbits shown in Fig.~\ref{fig:expolar}. Each system was integrated for 20\,000 yr ($\sim$ 650\,000 binary periods).}
\label{fig:polar.mass}
\vspace{-1.5em}
\end{center}
\end{figure}

Here we set the eccentricity of the binary to $e_B=0.5$ and vary the mass ratio $q$, and the mass of the planet $M_{\rm p}$ for polar orbits. By doing so we aim to explore whether the stability depends on the planetary mass. In Figure~\ref{fig:polar.mass}, we show the dynamical maps obtained for $\Delta \Omega$ and for $\Delta(\Omega-\varpi_{\rm B})$. For $q > 0.7$ and $M_{\rm p} < 10^{-5} \, M_{\odot}$, the node librates ($\Delta \Omega \sim 0^\circ$). For the same mass ratios but higher planetary masses we see that the node librates with higher amplitudes ($\Delta \Omega > 90^\circ$). Interestingly, for massive planets $M_{\rm p} > 10^{-5} \, M_{\odot}$ (i.e. giants) the orbit precesses regardless of the value of $q$. This occurs because the massive planet causes the binary to start precessing. In addition, we note that the value of $\Delta(\Omega-\varpi_{\rm B})$ does not depend on the planetary mass. Hence, in this configuration the indicator $\Delta \Omega$ allows us to discern whether the binary ellipse is fixed in time or not. Figure~\ref{fig:polar.mass} (lower frame) shows that, as expected, both the orbits of a single giant planet and of an almost massless planet remain polar around the osculating binary ellipse (i.e. independently of the binary precession).

\begin{figure}
\begin{center}
\includegraphics[width=0.45\textwidth]{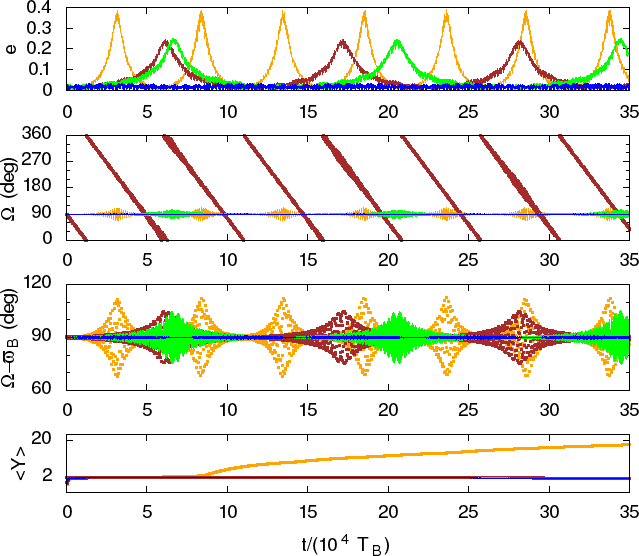}\\
\caption{Evolution of the individual polar orbits of Fig.~\ref{fig:polar.mass}: (1) blue, (2) green, (3) brown, and (4) yellow. The secular coupling between eccentricity, inclination, and node is easily identified. The periods are of roughly $5 \times 10^4$ binary periods, depending on the couple of values ($M_P$, $q$). The evolution of giant planets ($M_P > 10^{-4}$, i.e. brown region in Fig. \ref{fig:expolar}) exhibits the circulation of node. According to the {\sc megno} indicator, orbits 1, 2, and 3 are regular ($\langle Y \rangle \sim 2$); instead, 4 is chaotic after $\sim$100\,000 orbital periods of the binary ($\langle Y \rangle > 2$).}
\label{fig:expolar}
\vspace{-1.5em}
\end{center}
\end{figure}

Finally, in Figure~\ref{fig:expolar} we show the evolution of four selected orbits identified with numbers in Fig.~\ref{fig:polar.mass}. Orbit (1) corresponds to an orbit of an almost massless ($10^{-15} \, M_\odot$, roughly the mass of a planetary embryo) around a binary with $q=0.73$. Orbit (1) is to be compared with orbit (2), which corresponds to an embryo as well, but orbiting around a binary with $q=0.67$. Orbit (2) exhibits secular excitation in $e$ and $\Omega$, and it is stable for at least $10^8$ binary orbits. Orbit (3) corresponds to a planet of mass $10^{-3} \, M_\odot$ ($\sim$ Jupiter mass) orbiting around a binary with $q=0.67$. We observe that the node circulates and that the secular excitations in $\Delta e$ and $\Delta i$ are of the same order as in orbit (2). Finally, orbit (4) corresponds to the orbit of a small moon ($10^{-11} \, M_\odot$) around a binary with $q=0.2$. In this case the excitation in $e$ reaches values of $0.4$, which may eventually trigger instability. In fact, according to the {\sc megno} indicator, orbits 1, 2, and 3 are regular ($\langle Y \rangle \sim 2$), while orbit 4 is chaotic ($\langle Y \rangle > 2$). Based on the varied behaviour of these characteristic orbits, we conclude that planets with high masses ($M_{\rm p} \gtrsim 1 \times 10^{-5} \, M_\odot$) cause the binary pericentre to precess, and that the most regular orbits are obtained around binaries with $q>0.7$.

In summary, the most favourable conditions to obtain stable polar orbits for circumbinary planets are reached, at moderate to high inclinations, around equal-mass components and eccentric binaries for which $\Omega \sim \pm 90\degr$. In addition, planets with masses above $10^{-5} \, M_\odot$ strongly affect the binary orbit. Assuming that such a massive planet forms in the CBD, its gravitational perturbations are likely to render the surrounding disc material unstable. The study of this effect is beyond the scope of this work and will be more thoughtfully explored in the future.

%%%%%%%%%%%%%%%%%%%%%%%%%%%%%%%%%%%%%%%%

\section{Discussion}
\label{sec:discussion}

As mentioned in Sect.~\ref{sec:flyby}, the specific initial conditions for polar alignment can either be reached after a stellar flyby \citep{Cuello+2019b} or through misaligned accretion within the molecular cloud \citep{Bate2018}. Therefore, it is reasonable to consider polar alignment of the CBD as a possible (and potentially likely) dynamical outcome \citep{Aly+2015, Zanazzi&Lai2018, Lubow&Martin2018}. The presence of the gaseous disc around the binary is crucial in order to viscously damp the nodal libration mechanism of the CBD (see Sect~\ref{sec:polaralignment}.). This is indeed a necessary condition for polar alignment, which should be distinguished from polar oscillations.

The final disc configuration, however, is highly sensitive to the binary parameters. For example, increasing the binary eccentricity $e_{\rm B}$ (for a fixed $\Omega$) widens the inclination range for which polar alignment is expected. Additionally, increasing the binary mass ratio $q$ decreases the oscillation period (cf. Eq.~16 in \citealt{Lubow&Martin2018}) and makes the disc behave less rigidly. This is precisely when symmetry breaking between prograde and retrograde configurations is expected to occur. Then, for high values of $q$ and $e_{\rm B}$, the CBD is more susceptible to breaking up with at least one of the two discs eventually becoming polar. 

The disc parameters are equally important as shown by ML2018: a higher disc viscosity leads to a faster damping of the nodal libration mechanism. Also, cool discs could be more easily broken by the binary torque. When the disc precession timescale is shorter than the sound crossing time, then the disc breaks at roughly 3 $a_{\rm B}$. Finally, the larger the disc, the longer the timescale for the oscillations, even though these oscillations are more rapidly damped.

In this context, the very recent detection of a gas-rich circumbinary disc in polar configuration by \cite{Kennedy+2019} is of particular interest. In other cases, the presence of a highly inclined disc --- perhaps in a polar configuration --- has been proposed to explain asymmetric illumination patterns \citep{Marino+2015, Min+2017, Facchini+2018}. In some cases, these shadows are able to trigger spiral formation in the outer disc \citep{Montesinos+2016,Montesinos&Cuello2018}, which are able to efficiently trap dust \citep{Cuello+2019a}. Therefore, a shadow-casting inner disc is expected to affect the process of planet formation in the outer disc. Interestingly, the presence of shadows and spirals in an outer CBD could provide an indirect way to infer the presence of an inner CBD in a polar configuration.

Then, provided planet formation is efficient in these systems, polar and misaligned circumbinary P-type planets should be common around mildly eccentric ($e\leq0.4$) equal-mass binaries ($q>0.7$). We recall that planet formation on polar orbits is more favourable when $\delta e$ and $\delta i$ are small since these conditions guarantee a small velocity dispersion. This kind of planet is stable at distances of a few times the binary semi-major axis ($a \gtrsim 3 a_B$). For $\Omega \sim 90\degr$, polar stability islands appear at even closer distances from the binary. These regions should be considered as sweet spots for polar Tatooine formation and evolution. 

Interestingly, planets with masses above $10^{-5}\,M_{\odot}$ significantly affect the binary orbit and exhibit fast node circulation (alternatively binary pericentre precession). Observational campaigns such as {\sc bebop} \citep{Martin+2019} provide, and will continue to provide, useful constraints on their potential existence and frequency.

Last but not least, we also expect gaps at $1:N$ resonances with the binary once the gas dissipates in the remaining circumbinary debris disc. In the absence of any viscous force, these gaps naturally appear for high values of $\langle Y \rangle$ in Fig.~\ref{fig:ai}. This effect has profound implications for planetesimal growth and planet evolution in more evolved CBDs, which will be investigated in a future work. 

%%%%%%%%%%%%%%%%%%%%%%%%%%%%%%%%%%%%%%%%

\section{Conclusions}
\label{sec:conclusions}

We first studied the evolution of misaligned circumbinary discs through hydrodynamical simulations (see Section~\ref{sec:CBDs}). Then, by means of $N$-body integrations, we thoroughly characterised the dynamics and stability of single P-type planets around eccentric binaries (see Section~\ref{sec:characterisation}). The main results of this study can be summarised as follows:
\begin{enumerate}
\item The viscous torque exerted by the binary on the misaligned circumbinary disc (CBD) is stronger for retrograde configurations compared to prograde ones. Hence, for the same relative inclination with respect to $i=90\degr$, a retrograde CBD is more likely to become polar compared to a prograde CBD. This kind of symmetry breaking can happen in the inner disc regions under specific initial conditions (e.g. our run~3). Due to tilt and twist oscillations, the polar alignment condition of the CBD (initially not fulfilled) can be reached after a few hundred orbits. When this happens, the CBD stops behaving as a solid body (see Figs.~\ref{fig:r3-xyz} and \ref{fig:r3tilt}) and the disc breaks. This phenomenon is not captured by the linear theory and is more likely to occur around eccentric equal-mass binaries.
\item Our N-body simulations show that although `exotic' polar circumbinary P-type planets are stable on long timescales, namely over $10^7$ binary orbital periods. Based on stability criteria, we expect to find this kind of planet around binaries with mild eccentricities ($e_{\rm B}<0.4$) and for any value of the mass ratio $q$ (see Fig.~\ref{fig:ebJ}).
\item We find that above a certain planetary mass threshold (around $10^{-5}\,M_\odot$ in Fig.~\ref{fig:polar.mass}), the massive polar Tatooine strongly affects the binary orbit, which then starts precessing. The impact of the binary precession and the gravitational perturbations of such a giant planet on a hypothetical disc of planetesimals will be the subject of a future work.
\end{enumerate}

In conclusion, planet dynamics around misaligned and polar circumbinary discs is dramatically different compared to similar coplanar configurations. We note that a lack of detections of polar circumbinary planets, mainly because of observational biases, does not constitute proof of non-existence. Future radial-velocity and transit surveys around binaries will help to constrain the fraction of polar, inclined, and retrograde P-type planets. Equal-mass binaries with eccentricities $\sim 0.4$ should be considered as the most promising targets to discover polar Tatooines. Thus, since HD~98800BaBe has a high eccentricity ($e_{\rm B}=0.785\pm0.005$, \citealt{Kennedy+2019}), this system does not constitute a good candidate for harbouring such a planet.

%%%%%%%%%%%%%%%%%%%%%%%%%%%%%%%%%%%%%%%%

\section*{Acknowledgements}
We thank the anonymous referee and the editor for very valuable comments and suggestions that have significantly improved our work. We also thank Cristian Beaug\'e, Agnese Da Rocha Rolo, and Antoine Rocher for useful discussions during this project. NC acknowledges financial support provided by FONDECYT grant 3170680. The authors acknowledge support from CONICYT project Basal AFB-170002. The Geryon cluster at the Centro de Astro-Ingenieria UC was extensively used for the SPH calculations performed in this paper. BASAL CATA PFB-06, the Anillo ACT-86, FONDEQUIP AIC-57, and QUIMAL 130008 provided funding for several improvements to the Geryon cluster. N-body computations were performed at Mulatona Cluster from CCAD-UNC, which is part of SNCAD-MinCyT, Argentina. This project has received funding from the European Union's Horizon 2020 research and innovation programme under the Marie Sk\l{}odowska-Curie grant agreement No 823823. Figure~\ref{fig:r3-xyz} was made with {\sc splash} \citep{Price2007}.

\bibliographystyle{aa}
\bibliography{polar}

%%%%%%%%%%%%%%%%%%%%%%%%%%%%%%%%%%%%%%%%

\label{lastpage}
\end{document}